\documentclass[10pt,letterpaper]{article}
\usepackage{opex3}

\usepackage[utf8]{inputenc}
\usepackage{amsmath}

\input{colordvi.sty} 

\begin{document}

\title{Spatial and spectral properties of the pulsed second-harmonic generation
in a PP-KTP waveguide}

\author{%
    Radek Machulka$^{1,*}$, %
    Ji\v{r}\'{i} Svozil\'{i}k$^{1,2}$, %
    Jan Soubusta$^3$, %
    Jan Pe\v{r}ina, Jr.$^3$, %
    and Ond\v{r}ej Haderka$^1$%
}

\address{%
    $^1$RCPTM, Joint Laboratory of Optics PU and IP AS CR, 17. listopadu 12, 771~46 Olomouc, Czech Republic \\%
    $^2$ICFO---Institut de Ciencies Fotoniques, Mediterranean Technology Park, 08860, Castelldefels, Barcelona, Spain \\%
    $^3$Institute of Physics of Academy of Science of the Czech Republic,
        Joint Laboratory of Optics of PU and IP AS CR, 17. listopadu 12, 772~07 Olomouc, Czech Republic%
}

\email{$^{*}$radek.machulka@jointlab.upol.cz}

\begin{abstract}
Spatial and spectral properties of the pulsed second harmonic
generation in a periodically-poled KTP waveguide exploiting
simultaneously the first, second, and third harmonics of periodic
nonlinear modulation are analyzed. Experimental results are
interpreted using a model based on finite elements method.
Correlations between spatial and spectral properties of the
fundamental and second-harmonic fields are revealed. Individual
nonlinear processes can be exploited combining spatial and
spectral filtering. Also the influence of waveguide parameters to
the second-harmonic spectra is addressed.
\end{abstract}


\section{Introduction}

The process of second harmonic generation (SHG) has been one of
the most frequently studied nonlinear optical phenomena since the
invention of lasers. This process was observed by Franken et al.
in 1961 for the first time using crystalline quartz
\cite{Franken61}. In principle, a second-harmonic (SH) field can
be generated in many nonlinear materials that lack the inversion
symmetry. However, phase matching of the nonlinearly interacting
fields cannot be reached provided that the normal dispersion
occurs \cite{Helmy11}. Here, nonlinear anisotropic crystals with
their birefringence have been found useful and allowed to fulfil
phase-matching conditions.

Alternatively, nonzero phase mismatch $\Delta {\bf k}$ among the
interacting fields can be compensated by a periodic spatial
modulation of the nonlinear susceptibility with period $\Lambda_{\rm PM}$ ($\Lambda_{\rm PM} = 2\pi/|\Delta {\bf k}|$), as
has been suggested by Armstrong \cite{Armstrong62}. A commonly
used technique in this case is based on inverting dielectric
domains inside a ferro-electric nonlinear crystal using an intense
static electric field \cite{Fejer89,Yamada93}. This process is
called periodic poling (PP) and has allowed to utilize the highest
material nonlinear coefficients \cite{Fejer92}.

The poling process leaves well defined domains of the alternating
sings inside the nonlinear material. $\chi^{(2)} $ nonlinearity
can thus be approximated by a dichotomic function that can have
significant coefficients at its first- as well as higher-order
harmonics. Usually, a nonlinear structure is designed such that
only one harmonic is exploited. Contributions from other
harmonics are considered as parasitic \cite{Yu2005} in this case.
If domain lengths are chosen such that several harmonics of
nonlinear modulation are significant a rich spatial and spectral
modal structure in a waveguide can be exploited \cite{Chen2009}.
This gives a substantial increase in the ability to tailor
properties of the nonlinear process inside a waveguide. Here, we
study these abilities considering a waveguide that uses the first
three harmonics of nonlinear modulation.

Efficiency of the SHG process is inversely proportional to the
transverse area of the fundamental beam and, as a consequence,
beam focusing may improve the efficiency of SHG. However, walk-off
of the interacting fields puts limits to this property in bulk
crystals. For this reason, waveguiding structures not suffering
from this problem have become very important \cite{Roelofs94}.
They have allowed to confine fields' energies into a very small
area (typically of the order of tens of~$\mu m^{2}$) that results
in the desirable high power densities \cite{Kaleva08}. Intense SH
fields can then be obtained which, among others, allow to generate
squeezed light \cite{Perina1991,PerinaJr2007}.

Usually, waveguides support the propagation of the fundamental as
well as several higher-order modes, for which intensity
oscillations in the transverse plane are typical. The guided modes
differ in their wave vectors (propagation constants). The
phase-matching conditions and efficient SHG process can be reached
simultaneously for several combinations of these spatial modes
\cite{Karpinski09,Christ09,Mosley09}. For each of these individual
phase-matched processes, certain spectral and also spatial
characteristics of the fundamental beam are needed. We note that
as we are interested in pulsed SHG we consider spectrally
polychromanic modes determined by phase-matching conditions for
spatially defined modes. Different spectral properties for
different individual processes are useful in tuning a specific
nonlinear process. We note that, on the top of phase-matching
conditions, also a sufficient spatial overlap of electric-field
amplitudes of the interacting guided modes is needed. This
disqualifies higher-order modes whose amplitudes change their
signs several times in the transverse plane. Precise determination
of efficient combinations of the modes is thus very important in
characterizing the analyzed waveguide. This characterization is
useful not only for the SHG process. Also the inverse process of
spontaneous parametric down-conversion well known for its ability
to generate entangled photon pairs
\cite{Christ09,Mosley09,Banazsek01} benefits from the analysis.

Waveguides are promising not only as efficient media for nonlinear
interactions, they also allow for miniaturization and integration
into more complex circuits. Their nonlinearities allow for the
generation of squeezed light that is useful for optical
communications and quantum metrology \cite{Mandel1995}.

The theory of SHG in periodically poled waveguides is introduced
in Sec.~2. Equations giving the transverse profiles of the
fundamental and SH modes are derived in Sec.~3. The calculated and
measured SH spectra are analyzed in Sec.~4. Spectral and spatial
properties of type~II SHG are discussed in Sec.~5. Conclusions are
drawn in Sec.~6.

\section{Second-harmonic generation}

The evolution of electric-field amplitude ${\bf E}^{(2)}$ of the
SH field inside the waveguide is governed by the nonlinear wave
equation \cite{Mandel1995,Boyd08}:
\begin{equation} 
    \nabla\times\left(\nabla\times {\bf E}^{(2)}\right) +
  \frac{\overleftrightarrow{\varepsilon}}{c^{2}}\frac{\partial^{2}
   {\bf E}^{(2)}}{\partial t^{2}} =
  - \mu_{0}\frac{\partial^2{\bf P}^{(2)}}{\partial t^{2}};
\label{eq:wave}
\end{equation}
$\nabla = {\bf x}\frac{\partial}{\partial x} +{\bf
y}\frac{\partial}{\partial y} + {\bf z}\frac{\partial}{\partial
z}$. Linear permittivity tensor $\overleftrightarrow{\varepsilon}$
is assumed to have a diagonal form in the used cartesian
coordinate system $\{x,y,z\}$. Constant $c$ means speed of light
in vacuum and $\mu_{0}$ is vacuum permeability. Vector ${\bf
P}^{(2)}$ of the second-order polarization introduced in
Eq.~(\ref{eq:wave}) can be expressed as ${\bf P}^{(2)} =
2\varepsilon_{0}\overleftrightarrow{d}:{\bf E}^{(1)} {\bf
E}^{(1)}$, where $\overleftrightarrow{d}$ is the third-order
tensor of nonlinear coefficients and ${\bf E}^{(1)}$ denotes an
electric-field amplitude of the fundamental pumping field. Symbol
$: $ stands for tensor shortening with respect to its indices.

The electric field amplitudes of both the fundamental [${\bf
E}^{(1)}$] and SH [${\bf E}^{(2)}$] fields can be spectrally
decomposed as follows:
\begin{align} 
 {\bf E}^{(j)}(x,y,z,t) = \sum_{p={\rm TE,TM}}\sum_{n}\int d\omega
  {\cal E}^{(j)}_{pn}(z,\omega) {\bf e}^{(j)}_{pn}(x,y,\omega)
  \exp{\left[i\beta^{(j)}_{pn}(\omega)z
  - i\omega t\right]}, \hspace{3mm} j=1,2.
\label{eq:cplxAmpl}
\end{align}
The normalized electric-field mode functions ${\bf
e}_{pn}^{(j)}(x,y,\omega) $ form a basis and are obtained as
eigenfunctions of the Helmholtz equation (see later in
Sec.~\ref{sec_mody}). The Helmholtz equation also provides the
corresponding propagation constants $\beta^{(j)}_{pn}(\omega) $.
A detailed analysis of modal properties of the studied waveguide
has shown that two groups of modes differing in polarization exist
\cite{Snyder83}. Whereas (quasi-) TE modes have electric-field
polarization vectors nearly parallel to the $x$ axis,
electric-field polarization vectors of (quasi-) TM modes are
nearly perpendicular to the $x$ axis. In addition, there exists
a system of transverse modes labelled by index $n$. Finally,
envelopes ${\cal E}_{pn}^{(j)}(z,\omega)$ of the spectral
amplitudes of the corresponding modes propagating along the $+z$
axis have been introduced in Eq.~(\ref{eq:cplxAmpl}).

Substituting the decomposition of electric-field amplitudes ${\bf
E}^{(j)}$ in Eq.~({\ref{eq:cplxAmpl}) into the wave equation
(\ref{eq:wave}) a set of nonlinear differential equations for the
envelope amplitudes ${\cal E}_{pn}^{(j)}(z,\omega)$ can be
derived. Invoking the slowly varying amplitude approximation and
non-depleted fundamental pump field approximation, the following
formula for envelope amplitude ${\cal E}^{(2)}_{ak}$ at the end
of the waveguide of length $L$ has been obtained after simple
integration:
\begin{eqnarray} 
 {\cal E}^{(2)}_{ak}(L,\omega_{2}) &=&
    \frac{i\omega^{2}_{2}}{c^{2}\left[\beta^{(2)}_{ak}(\omega_{2})\right]^{2}}
    \sum_{b,c={\rm TE,TM}} \,\, \sum_{l,m} \int d\omega_{1} D^{abc}_{klm}(\omega_{2},\omega_{1})
     \Gamma^{abc}_{klm}(L,\omega_{2},\omega_{1}) \nonumber \\
 & & \hspace{1cm} \mbox{} \times {\cal E}^{(1)}_{bl}(0,\omega_{1})
     {\cal E}^{(1)}_{cm}(0,\omega_{2} - \omega_{1}) .
\label{eq:scf}
\end{eqnarray}
We note that the integration over frequency $\omega_1$
incorporates all frequency contributions occurring in the
polychromatic fundamental field. In Eq.~(\ref{eq:scf}), effective
nonlinear coefficients $ D^{abc}_{klm} $ have been introduced:
\begin{equation} 
    D^{abc}_{klm}(\omega_{2},\omega_{1}) = d_M
    \int_{-\infty}^{\infty} dx \int_{-\infty}^{\infty} dy
     \overleftrightarrow{d} : {\bf e}^{(2)*}_{ak}(x,y,\omega_{2})
      {\bf e}^{(1)}_{bl}(x,y,\omega_{1})
      {\bf e}^{(1)}_{cm}(x,y,\omega_{2} - \omega_{1}).
\label{eq:D}
\end{equation}
The effective nonlinear coefficients $ D^{abc}_{klm} $ incorporate
in their definitions both polarization properties and overlap of
the transverse parts of the interacting modes amplitudes. Constant
$ d_M $ gives the amplitude of decomposition of the actual
modulation of nonlinear coefficient into the $ M $-th harmonic.
The coefficients $ D^{abc}_{klm} $ are assumed to be weakly
frequency dependent. Considering nonlinear coefficients $
D^{abc}_{klm} $ involving higher-order spatial modes, their values
are usually small due to frequent changes in the sign of the
electric-field amplitudes of these modes in the transverse plane.
The coupling functions $\Gamma^{abc}_{klm}(L,\omega_2,\omega_1) $
introduced in Eq.~(\ref{eq:scf}) characterize fields' evolution
along the $ z $ axis:
\begin{equation} 
    \Gamma^{abc}_{klm}(L,\omega_{2},\omega_{1}) = i\frac{
     \exp{\left[-i\Delta\beta^{abc}_{klm}\left(\omega_{2},\omega_{1}\right)L\right]}
     - 1 }{ \Delta\beta^{abc}_{klm}\left(\omega_{2},\omega_{1} \right) } .
\label{eq:cf}
\end{equation}
The nonlinear phase mismatch $\Delta\beta^{abc}_{klm}$ occurring
in Eq.~(\ref{eq:cf}) is defined as
\begin{equation} 
 \Delta\beta^{abc}_{klm}\left(\omega_{2},\omega_{1}\right) =
    \beta^{(2)}_{ak}(\omega_{2})
    - \beta^{(1)}_{bl}\left(\omega_{1}\right)
    - \beta^{(1)}_{cm}\left(\omega_{2} - \omega_{1}\right)
    - \frac{2\pi M}{\Lambda_{\rm PM}};
\label{eq:pmf}
\end{equation}
the last contribution $2 \pi M/\Lambda_{\rm PM}$ originates in the
$M$-th harmonic of the periodic modulation of nonlinear
coefficient.

\section{Equations giving fields' spatial profiles}
\label{sec_mody}

Material of the waveguide, KTP, is anisotropic and is positioned
such that only diagonal elements $ \varepsilon_x $, $
\varepsilon_y $, and $ \varepsilon_z $ of the linear
susceptibility tensor are nonzero. The crystallographic axes of
KTP ($x_c$, $y_c$, $z_c$) coincide with the axes of the coordinate
system ($z$, $x$, $y$). The elements $\varepsilon_x$,
$\varepsilon_y$, and $\varepsilon_z$ can be expressed using
indices of refraction $n_{\xi}$ ($\varepsilon_{\xi} = n_{\xi}^{2}
$), $\xi=x,y,z$, that take the following form in the studied
waveguide \cite{Roelofs94,Fan87,Bierlein87}:
\begin{eqnarray} 
  \left. n_{\xi}(x,y) \right|_{y \geq 0}  &=& n_{\xi 0} + \Delta n_{\xi}
   \mathrm{rect}_{[-w/2,w/2]}(x)\, \mathrm{erfc}\left(-y/h \right), \nonumber \\
  \left. n_{\xi}(x,y) \right|_{y < 0} &=& 1 ; \hspace{2cm} \xi = x,y,z.
\label{eq:index}
\end{eqnarray}
The indices of refraction $ n_{\xi 0} $ characterize the
substrate. Changes $\Delta n_{\xi} $ of the indices of refraction
are introduced in the process of waveguide formation. For
$\lambda_{1}$ = 800~nm ($\lambda_{2}$ = 400~nm) the following
values have been used: $n_{x} = 1.75719$, $\Delta n_{x} = 0.009$,
$n_{y} = 1.84546$, $\Delta n_{y} = 0.013$ ($n_{x} = 1.84435$,
$\Delta n_{x} = 0.018$, $n_{y} = 1.96775$, $\Delta n_{y} = 0.019$)
\cite{Fan87}. Above the waveguide, air is assumed. Waveguide's
depth is characterized by parameter $h$, whereas parameter $w$
denotes waveguide's width. We have assumed $h = 10~\mu$m and $w =
5~\mu$m for the analyzed waveguide. These values have been checked
experimentally by measuring the profile of white light leaving the
waveguide and comparing the obtained profile with that arising
from the model (see Fig.~\ref{fig:wlc}). The rectangular function
$ {\rm rect}_{[-w/2,w/2]}(x) $ equals 1 in the interval
$\langle-w/2,w/2\rangle $ and is zero otherwise. Also the error
function $ {\rm erfc}(x) $ has been invoked in
Eq.~(\ref{eq:index}).
\begin{figure} 
\centering
    \includegraphics[width=0.7\textwidth]{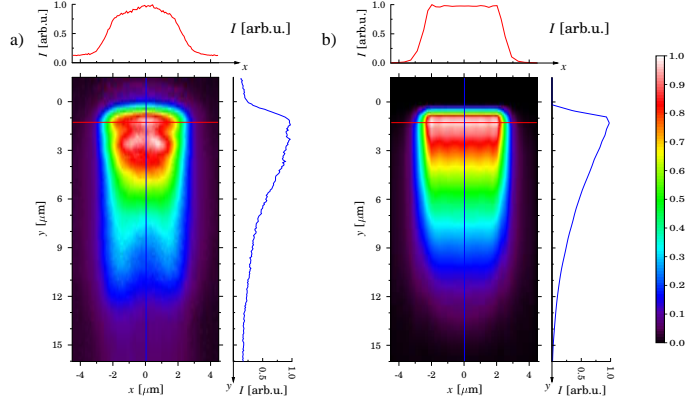}
\caption{Topo graphs of (a) experimental and (b) theoretical
intensity $I = |{\cal E}^{(1)}|^{2}$ in
the transverse plane formed by white light propagating inside the waveguide.
Also cuts along the lines indicated in the topo graphs are shown.}
\label{fig:wlc}
\end{figure}

Usually, TE modes are conveniently obtained by solving the wave
equation for the electric-field amplitude. The wave equation for
the magnetic-field amplitude is then useful in determining TM
modes. However, the spatial dependence of indices of refraction in
the studied waveguide causes difficulties and asymmetry in the
obtained equations. As a consequence, the use of wave equation for
the magnetic-field amplitude $ {\bf H} $,
\begin{equation} 
    \nabla\times\left( \overleftrightarrow{\varepsilon}^{-1} \nabla\times{\bf H}\right) =
    - \frac{1}{c^{2}}\frac{\partial^{2}{\bf H}}{\partial t^{2}},
\label{e2}
\end{equation}
is preferred in both cases.

The substitution $ {\bf H}(x,y,z,t) = {\bf h}(x,y,\omega)
\exp(i\beta z -i\omega t) $ in Eq.~(\ref{e2}) provides a set of
equations for cartesian components of the magnetic-field envelope
$ {\bf h}(x,y,\omega) $ of a mode characterized by frequency $
\omega $ and propagation constant $ \beta $. Considering the
relation
\begin{equation} 
 {\bf h}_z = i (\partial {\bf h}_x/\partial x +
 \partial {\bf h}_y/\partial y) / \beta
\label{eq:hz}
\end{equation}
originating in the Maxwell equation $ \nabla {\bf H} = 0 $, two
coupled differential equations for the components $ {\bf h}_x $
and $ {\bf h}_y $ can be derived:
\begin{eqnarray} 
 \frac{1}{\varepsilon_{y}}\frac{\partial^{2}{\bf h}_{x}}{\partial x^{2}} +
  \frac{1}{\varepsilon_{z}}\frac{\partial^{2}{\bf h}_{x}}{\partial y^{2}} +
  \frac{1}{\varepsilon_{y}}\frac{\partial^{2}{\bf h}_{y}}{\partial x\partial y} -
  \frac{1}{\varepsilon_{z}}\frac{\partial^{2}{\bf h}_{y}}{\partial y\partial x}
  - \left[\frac{\partial {\bf h}_{y}}{\partial x} -
  \frac{\partial {\bf h}_{x}}{\partial y}\right]
  \frac{\partial}{\partial y}\left(\frac{1}{\varepsilon_{z}}\right)
  = \left(\frac{\beta^{2}}{\varepsilon_{y}} - k^{2}_{0}\right){\bf h}_{x},
  \label{eq:svfHx}   \\
 \frac{1}{\varepsilon_{z}}\frac{\partial^{2}{\bf h}_{y}}{\partial x^{2}} +
  \frac{1}{\varepsilon_{x}}\frac{\partial^{2}{\bf h}_{y}}{\partial y^{2}} +
  \frac{1}{\varepsilon_{x}}\frac{\partial^{2}{\bf h}_{x}}{\partial y\partial x} -
  \frac{1}{\varepsilon_{z}}\frac{\partial^{2}{\bf h}_{x}}{\partial x\partial y}
  + \left[\frac{\partial {\bf h}_{y}}{\partial x} -
  \frac{\partial{\bf h}_{x}}{\partial y}\right]
  \frac{\partial}{\partial x}\left(\frac{1}{\varepsilon_{z}}\right)
  = \left(\frac{\beta^{2}}{\varepsilon_{x}} - k^{2}_{0}\right){\bf h}_{y},
\label{eq:svfHy}
\end{eqnarray}
$k_{0} = \omega/c$. The last component $ {\bf h}_z(x,y,\omega) $
is then obtained from Eq.~(\ref{eq:hz}) provided that the
remaining components have been determined. Knowing the
magnetic-field envelope $ {\bf h} $, the electric-field envelope $
{\bf e} $ is derived from the Maxwell equation that gives $ {\bf
e}(x,y,\omega) = \overleftrightarrow{\varepsilon}^{-1} [i\nabla
\times {\bf h}(x,y,\omega) - \beta {\bf z} \times {\bf
h}(x,y,\omega) ] / (\varepsilon_0 \omega ) $; $ {\bf z} $ means
the unit vector along the $ z $ axis.

Considering Eqs.~(\ref{eq:svfHx}) and (\ref{eq:svfHy}), the cross
contributions
\begin{align}
  \frac{1}{\varepsilon_{y}}\frac{\partial^{2}{\bf h}_{y}}{\partial x\partial y},\quad
  \frac{1}{\varepsilon_{z}}\frac{\partial^{2}{\bf h}_{y}}{\partial y\partial x},\quad &
  \left[\frac{\partial {\bf h}_{y}}{\partial x} - \frac{\partial {\bf h}_{x}}{\partial y}\right]
    \frac{\partial}{\partial y}\left(\frac{1}{\varepsilon_{z}}\right),
\nonumber \\
  \frac{1}{\varepsilon_{x}}\frac{\partial^{2}{\bf h}_{x}}{\partial y\partial x},\quad
  \frac{1}{\varepsilon_{z}}\frac{\partial^{2}{\bf h}_{x}}{\partial x\partial y},\quad &
  \left[\frac{\partial {\bf h}_{y}}{\partial x} - \frac{\partial {\bf h}_{x}}{\partial y}\right]
    \frac{\partial}{\partial x}\left(\frac{1}{\varepsilon_{z}}\right)
\nonumber
\end{align}
are usually very small and so they can be omitted \cite{Kawano01}.
As a result, equations (\ref{eq:svfHx}) and (\ref{eq:svfHy})
become independent and so can be solved separately giving (quasi-)
TE and TM modes. This considerably simplifies the analysis. The
obtained equations can be written as follows:
\begin{eqnarray} 
 \frac{\partial^{2}{\bf h}_{x}}{\partial x^{2}} + \frac{\varepsilon_{y}}{{\varepsilon}_{z}}
  \frac{\partial^{2}{\bf h}_{x}}{\partial y^{2}} &=& \left(\beta^{2} - \varepsilon_{y} k^{2}_{0}\right) {\bf h}_{x},
  \nonumber  \\
 \frac{\varepsilon_{x}}{\varepsilon_{z}} \frac{\partial^{2}{\bf h}_{y}}{\partial x^{2}} +
  \frac{\partial^{2}{\bf h}_{y}}{\partial y^{2}} &=& \left(\beta^{2}
  - \varepsilon_{x} k^{2}_{0}\right) {\bf h}_{y}.
\label{eq:hx}
\end{eqnarray}

For certain profiles of permittivity
$\overleftrightarrow{\varepsilon}$, the solution of
Eqs.~(\ref{eq:hx}) can be found analytically \cite{Snyder83,Rubinstein1998}. For
example, the simplest approximation of waveguide's profile by the
rectangular functions in both $ x $ and $ y $ directions together
with the known analytical form of eigenmodes has been used in
\cite{Christ09} to interpret the experimental results. However,
only the numerical approach is possible for transverse profiles of
real waveguides. For this reason, a large variety of sophisticated
numerical methods has been developed \cite{Kawano01}. Among them,
a finite difference method has become popular because of simple
implementation, even in its full-vector form. Unfortunately, it
requires a large amount of computer resources for solving the
eigenvalue problem. This represents a problem when more complex
transverse profiles are considered. Nevertheless, it has been
successfully applied both in its scalar \cite{Fiorentino2007} as
well as semi-vectorial \cite{Karpinski09,Fallahkhair2008} forms.

On the other hand, a finite elements method
\cite{Kawano01,Jin02} has been found more suitable due to
its stability and lower demands on computer resources. It is
resistant against non-physical solutions. Moreover, the use of
higher-order approximation functions results in lowering of the
dimension of eigenvalue problem. For this reason, many commercial
waveguide mode-solvers are based upon this method. For instance,
it has been applied for the determination of waveguide's mode
dispersion in \cite{Spillane2007}. In our investigation, we have
developed an implementation of the finite elements method that is
based on the Galerkin method \cite{Kawano01}. Easy
parallelization of this implementation makes it superior above
other approaches.

Intensity profiles of the first three modes both for the
fundamental and SH TE-polarized fields as they emerge from the
developed method are depicted in Fig.~\ref{fig:modes}. The
transverse profiles of TM-polarized modes are similar to those
appropriate to the corresponding TE-polarized modes. According to
the convention, spatial modes are labelled by two numbers that
give the number of nodes along the $x$ and $y$ axes. The numerical
analysis has revealed that the analyzed waveguide supports more
than 30 spatial modes at $\lambda_{2} = 400$~nm for both TE and TM
polarizations and 3 (5) spatial modes at $\lambda_{1} = 800$~nm
for TE (TM) polarization.

\begin{figure} 
    \includegraphics[width=\textwidth]{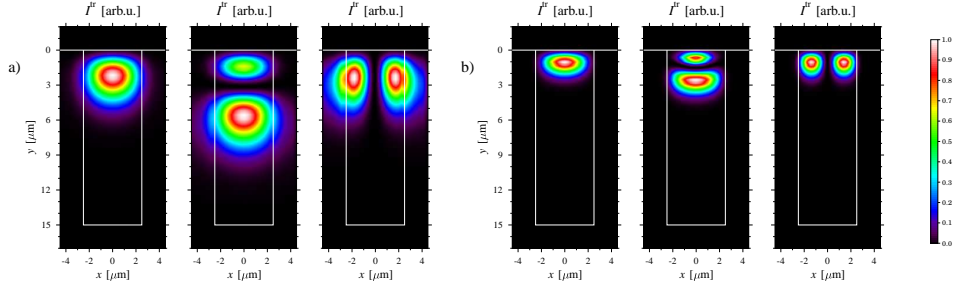}
\caption{Topo graphs of calculated electric-field intensity distributions
$ I^{\rm tr} $ ($I^{\rm tr}(x,y)=|{\bf e}(x,y)|^2$) in the
transverse plane for the first three spatial modes
denoted as (0,0), (0,1), and (1,0) for the fundamental TE-polarized (a)
and SH TE-polarized (b) field;
$ w= 5 $~$\mu$m, $ h = 10 $~$\mu$m. Frame boxes in the graphs
indicate boundaries of the waveguide.}
\label{fig:modes}
\end{figure}

\section{Experimental second harmonic generation in the waveguide}

The analyzed waveguide was fabricated on a KTP substrate ($ 10.5
\times 2 \times 1 $~mm$^3$) by Rb$^+$ ion diffusion together with
other cca 50 similar waveguides (manufacturer AdvR Inc.). The
requirement for an efficient type~II SHG at $\lambda_{1} = 800$~nm
has resulted in poling the nonlinear material with period
$\Lambda_{\rm PM} = 7.62~\mu$m. In the fabrication process both
horizontal and vertical dimensions of the waveguide are under
control. The horizontal width is given by geometry of the
lithographic mask that leads to a profile with well defined
boundaries owing to highly anisotropic diffusion. On the other
hand, the profile in the vertical direction is formed by ion
diffusion gradually in time. At some instant the profile of
refractive index becomes temporally stable and can be approximated
by the error function \cite{Bierlein87}.

Due to imperfections in the fabrication process, fluctuations in
waveguide's parameters including its width $w$, depth $h$, duty
ratio of the poling period as well as the poling period itself
inevitably occur. Moreover, there exist several areas without
poling inside the structure and also the poling duty ratio may
change from 50~\% to 75~\%. These fluctuations result in
broadening of the SH spectrum. To judge the amount of spectral
broadening we have plotted the dependence of SH wavelength shift $
\Delta \lambda_2 $ on waveguide's width $ w $, depth $ h $ and
poling period $ \Lambda_{\rm PM} $ in Fig.~\ref{fig:var} for five
individual nonlinear processes dominating in SH spectrum.
According to the manufacturer \cite{Kaleva12}, the expected
variations in waveguide's width $ w $, depth $ h $ and poling
period $ \Lambda_{\rm PM}$ are in turn $\pm$0.1~$\mu$m,
$\pm$2~$\mu$m and $\pm$0.1~$\mu$m. This in accord with
the curves of Fig.~\ref{fig:var} results in the estimated spectral
SH widths equal approximately to 0.05~nm, 0.1~nm, and 2~nm,
respectively. Fluctuations in the poling period $ \Lambda_{\rm PM}
$ thus represent the main source of SH spectral broadening
observed with a sufficiently broad fundamental spectrum.
Imperfections in nonlinear modulation even lead to the occurrence
of three SH fields originating in different polarization
configurations and having different ideal values of poling period
$ \Lambda_{\rm PM} $ (see below).

Larger differences in the values of width $ w $, depth $ h $ and
poling period $ \Lambda_{\rm PM} $ lead to significant shifts of
spectral positions of individual lines that result in the change
of the overall SH spectral profile. The character of this change
is illustrated in Fig.~\ref{fig:spcHor} showing the calculated SH
intensity spectrum $I$ for three values of the width $w$.

\begin{figure}[h!] 
\centering
    \includegraphics[width=\textwidth]{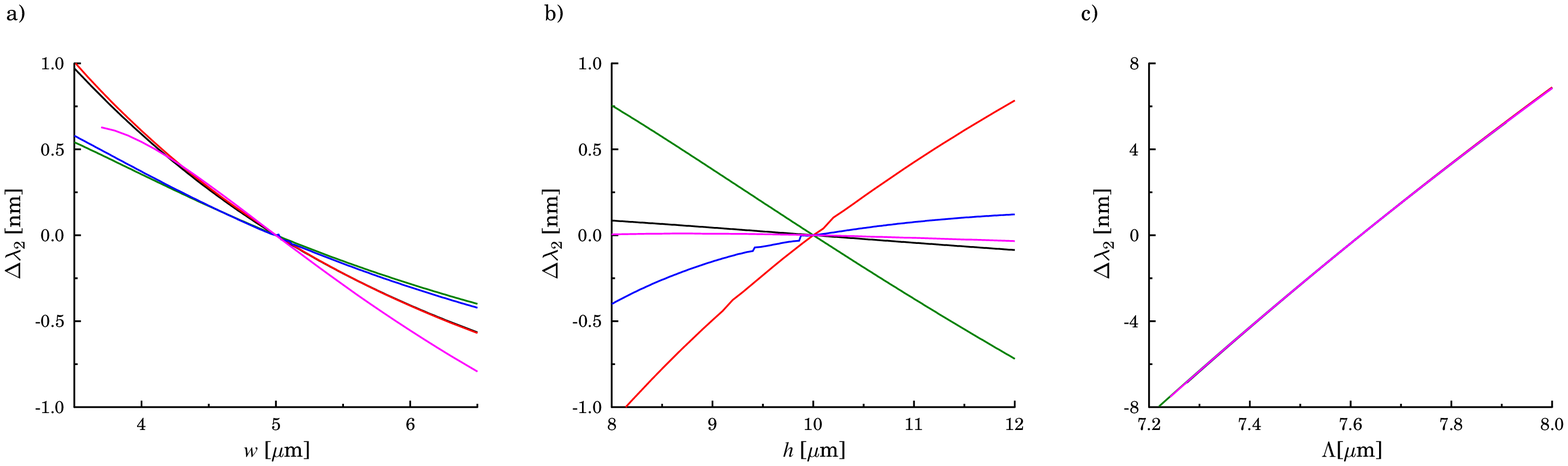}
 \caption{SH wavelength shift $ \Delta \lambda_2 $ as it depends on
 waveguide's width $ w $ (a), depth $ h $ (b) and poling period $ \Lambda_{\rm PM} $
 (c) for the following individual processes: (0,0) + (0,0) $\rightarrow$
(0,0) (black curve), (0,0) + (0,0) $\rightarrow$ (0,1) (red
curve), (0,1) + (0,1) $\rightarrow$ (0,0) (green curve), (0,1) +
(0,1) $\rightarrow$ (0,1) (blue curve), and (1,0) + (0,0)
$\rightarrow$ (1,0) (magenta curve). In (c), all curves nearly
coincide.} \label{fig:var}
\end{figure}

\begin{figure} 
\centering
    \includegraphics[width=\textwidth]{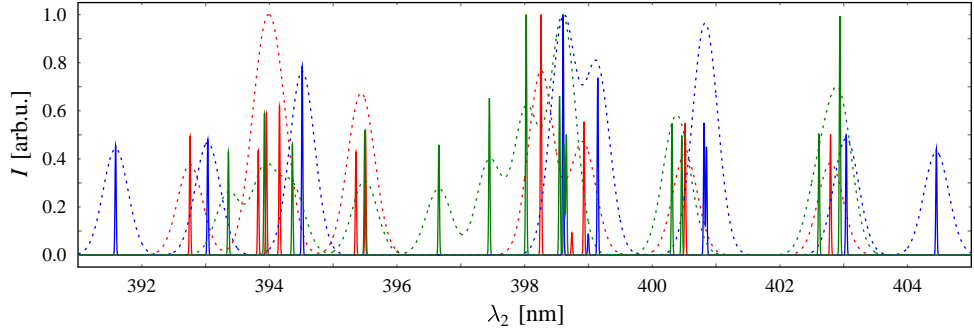}
\caption{SH spectral intensity $I$ depending on wavelength
$\lambda_{2}$ and calculated for different values of waveguide's
width $w$; $I(\omega_{2}) = \sum_{ak}|{\cal E}_{ak}
(L,\omega_{2})|^2 $. Solid curves correspond to ideal spectral
resolution 0.13~nm (given by phase-matching conditions along the
$z$ axis for a 10.5~mm long waveguide with the inclusion of
material and waveguiding dispersion). Dashed curves are
appropriate for the expected real spectral broadening of 1~nm; $w
= 4$~$\mu$m (blue curve), $w = 5$~$\mu$m (red curve) and $w =
6$~$\mu$m (green curve), $h = 10$~$\mu$m.} \label{fig:spcHor}
\end{figure}

To observe the full profile of a spectral line characterized by
its width $ \Delta \lambda_2 $, the width $ \Delta\lambda_1 $ of
the fundamental field has to be sufficiently large. Consideration
of energy conservation in the nonlinear process provides in this
case the relation $ \Delta \lambda_1 = 2\sqrt{2} \Delta \lambda_2
$. However, the needed width $ \Delta \lambda_1 $ has to be larger
than $ 2\sqrt{2} \Delta \lambda_2 $ because of phase-matching
conditions that put additional constraints to the nonlinear
process. This relation between the widths $ \Delta \lambda_1 $ and
$ \Delta \lambda_2 $ also determines how wide area in the SH
spectrum is observed for a given fundamental spectral width $
\Delta \lambda_1 $. We note that a more detailed information about
the phase-matching conditions and their spectral dependence inside
a broadened line can be reached by sum-frequency generation
utilizing spectrally narrow fields \cite{Karpinski09}.

The experimental investigation of SHG process has been carried out
in the setup sketched in Fig.~\ref{fig:setup}. Light from a
tunable femtosecond Ti-sapphire laser (87~MHz repetition rate,
100~fs pulse duration, central wavelengths in the range
790---810~nm, spectral width 10~nm) has been attenuated combining
half-wave plate ($\lambda/2$) and linear polarizer to attain the
power of about 20~mW. Another half-wave plate ($\lambda/2$) has
been used to control the pump-beam polarization. The pump-beam
central wavelength has been adjusted by tilting an interference
filter defining the spectral width of 3~nm. A small fraction of
the beam has been deflected to allow for precise monitoring of the
pump-beam position in the transverse plane needed for efficient
coupling of the beam into the waveguide using a 10$\times$
microscope objective. The light leaving the waveguide and composed
of both the pump and SH beams has been imaged by a 20$\times$
objective with a larger numerical aperture. A dichroic
beam-splitter has separated the pump beam from the SH beam. The
obtained beams have been analyzed by power-meter, CCD camera or
spectrometer (resolution 0.1~nm) after passing through a linear
polarizer.
\begin{figure} 
    \includegraphics[width=\textwidth]{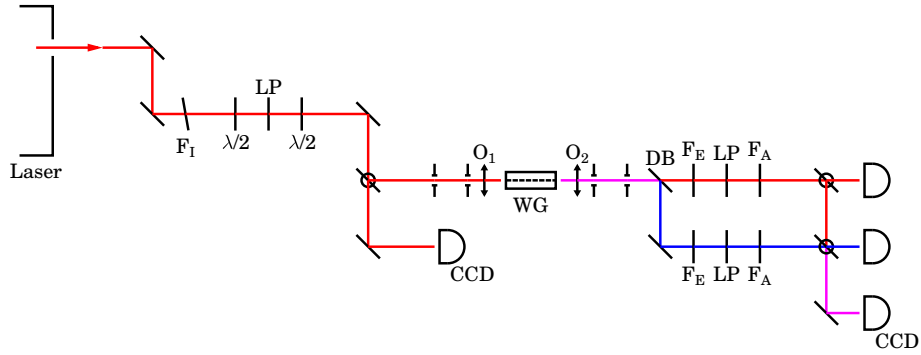}
\caption{Sketch of the experimental setup for waveguide (WG) investigation.
Symbol $\mathrm{F_{I}}$
denotes interference filter, $\mathrm{F_{A}}$ neutral density
filter, $\mathrm{F_{E}}$ edge-pass filter, $\lambda/2$ half-wave
plate, LP linear polarizer, O microscope objective, DB dichroic
beam-splitter, and CCD means CCD camera.}
\label{fig:setup}
\end{figure}
As the indices of refraction in KTP are temperature dependent, the
KTP chip has been thermally stabilized at 25\,{\textcelsius} using
a Peltier element with temperature sensor and feedback loop. We
note that the ideal phase-matching condition shifts by cca 1~nm
per 25\,{\textcelsius}.

For the chosen orientation of KTP, three types of SHG processes
differing in fields' polarization properties can occur
simultaneously due to different harmonics of the periodical
modulation of nonlinearity. In type~0 SHG, both the fundamental
and SH fields have TM polarizations. If the fundamental beam is TE
polarized and the SH beam TM polarized type~I SHG is observed. A
TE-polarized SH beam occurs only if both TE- and TM-polarized
components of the fundamental beam are present (type~II SHG).
Different conversion efficiencies are found for the discussed
processes because of different values of used nonlinear
coefficients, overlaps of mode functions, and also phase-matching
conditions. Especially, phase-matching conditions depending on the
value of poling period $\Lambda_{\rm PM}$ allow to control these
efficiencies. The analyzed waveguide was primarily optimized for
type~II SHG giving $\Lambda_{\rm PM} = 7.62~\mu$m. Nevertheless,
type~0 SHG (with the optimum poling period 3.08~$\mu$m) can be
observed exploiting the second harmonic of nonlinear periodic
modulation. The presence of second harmonic of nonlinear periodic
modulation together with all even harmonics is caused by the
declination of the actual nonlinear modulation from the ideal
shape with positive and negative domains of equal lengths
\cite{Yu1999}. We note that a nonlinear modulation with equal
lengths of positive and negative domains has only odd harmonics
nonzero. Also, the third harmonic of nonlinear periodic modulation
allows to approximately arrive at the phase-matching conditions of
type~I SHG for which the poling period 1.83~$\mu$m would have been
optimum. The experimental nonlinear conversion efficiencies $\eta$
in this configuration are summarized in Tab.~\ref{tab1}}. Type~II
SHG is the most efficient process due to optimum phase-matching
conditions and despite the lower value of nonlinear coefficient $
d_{32} $ compared to that of coefficient $ d_{33} $ used in type~0
SHG. Also better phase-matching conditions of type~I SHG allow
more efficient SHG than in type~0. However, and importantly, all
three types of SHG have comparable efficiencies. Their relative
values can suitably be changed by changing the poling period
$\Lambda_{\rm PM}$.
\begin{table} 
\centering
\begin{tabular}{l c c c}
    \hline
    \hline
    \multicolumn{2}{c}{Process} & $d$ \,[pm$\cdot$V$^{-1}$] & $\eta$\,[W$^{-1}\cdot$cm$^{-2}$] \\
    \hline
    type 0 & TM + TM $\rightarrow$ TM & $ d_{33}$ = 10.7 & 2.83 $\pm$ 0.06 \\
    type I & TE + TE $\rightarrow$ TM & $ d_{32}$ = 2.65 & 4.44 $\pm$ 0.12 \\
    type II & TE + TM $\rightarrow$ TE & $ d_{32}$ = 2.65 & 4.78 $\pm$ 0.03 \\
    \hline
\end{tabular}
\caption{Types of observed SHG processes, their polarizations,
used nonlinear coefficients $d$ \cite{Boyd08},
and experimental nonlinear conversion efficiencies $ \eta $ are given.
Conversion efficiency $ \eta $ has been determined as $
\eta = P_{\rm SHG} P_{\rm in}^{-2}L^{-2} $, where
$P_{\rm SHG}$ ($P_{\rm in}$) gives the power of outgoing SH
(coupled incident pump) beam and $L$ is the length of nonlinear
medium \cite{Kaleva08}.}
\label{tab1}
\end{table}

\begin{figure} 
    \includegraphics[width=\textwidth]{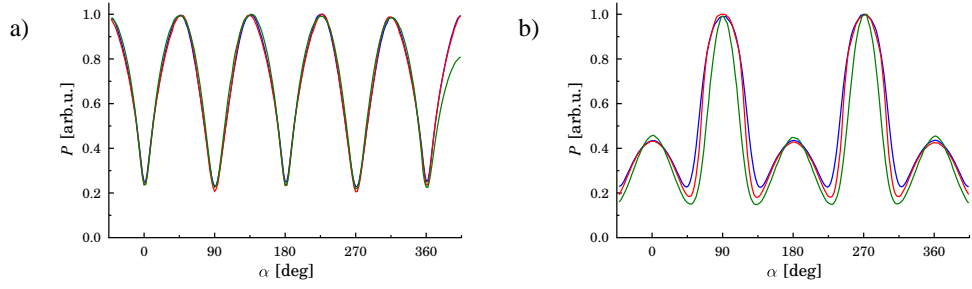}
\caption{Measured normalized power $P$ of SH field as it depends on angle
$\alpha$ giving the orientation of linear polarization of the
incident pump beam for (a) TE- and (b) TM-polarized SH field
($\alpha = 0$~deg corresponds to TE polarization).
Curves characterizing three
different waveguides are shown for comparison to judge fabrication capabilities.
The curves were obtained with experimental relative error 4~\%.}
\label{fig:pol}
\end{figure}

Suitable choice of polarization directions of both the fundamental
and SH fields allows to extract individual types of SHG. The
intensity of SH field in the most efficient type~II SHG process
can be separated from the overall field by using a polarization
analyzer aligned along TE polarization. The experimental evidence
of observing this process is given by measuring periodic
oscillations of SH intensity depending on the orientation of
linear polarization of the incident pump beam [see
Fig.~\ref{fig:pol}(a)]. Aligning the polarization analyzer in SH
field along TM polarization the remaining two processes are
visible. Changing the orientation of linear polarization of the
incident pump beam, we continuously move from one process to the
other [see Fig.~\ref{fig:pol}(b)]. Whereas TM polarization of the
pump beam supports type~0 SHG, TE polarization is suitable for
type~I SHG.

We further pay attention to the most efficient type~II SHG
process. The necessity to have both TE- and TM-polarized
fundamental field distinguishes it from the remaining two types.
This is important from the point of view of the reverse process of
parametric down-conversion
\cite{Christ09,Mosley09,Mandel1995,PerinaJr2008,Levine10} in which
pairs of photons with orthogonal polarizations are emitted. The
orthogonal polarizations of two photons would then allow to spatially
separate both photons using a polarizing beam-splitter.

\section{Spatial and spectral properties of type~II second harmonic generation}

\begin{figure} 
    \includegraphics[width=\textwidth]{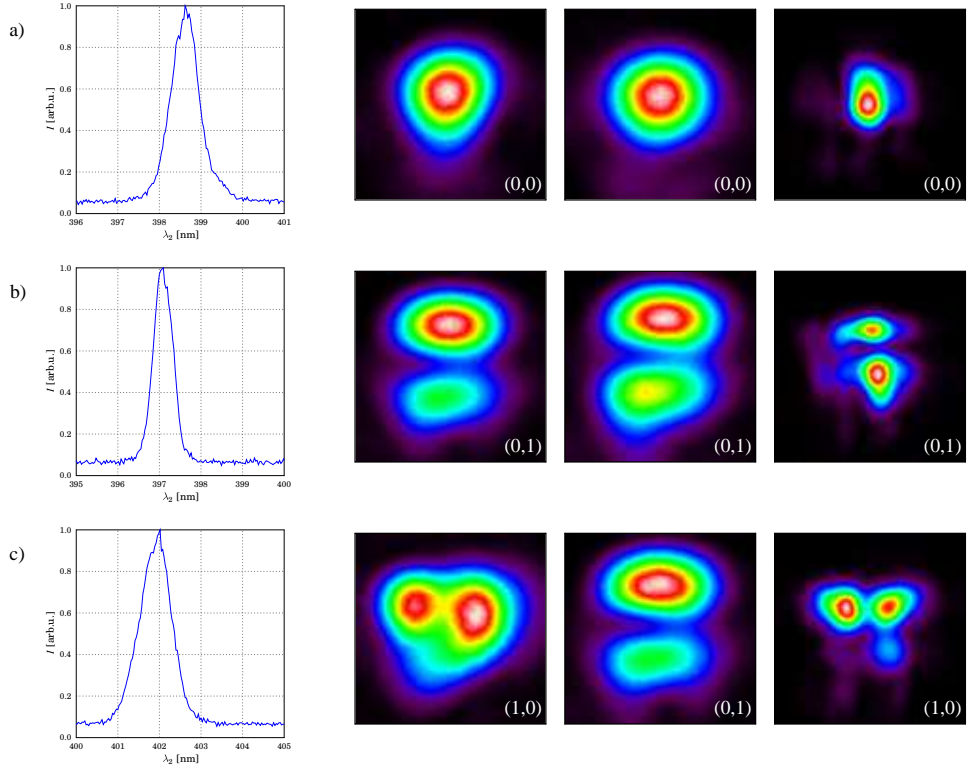}
\caption{Experimental SH spectral intensity $I$ depending on
wavelength $\lambda_{2}$ and topo graphs of spatial electric-field
intensity profiles $I^{\rm tr}(x,y)$ of TE-polarized pump beam,
TM-polarized pump beam, and TE-polarized SH beam are shown in
sequence for individual processes: (a) (0,0) + (0,0) $\rightarrow$
(0,0), (b) (0,1) + (0,1) $\rightarrow$ (0,1), and (c) (1,0) +
(0,1) $\rightarrow$ (1,0).} \label{fig:res}
\end{figure}

In agreement with the numerical analysis, the experiment has
revealed three (five) spatial TE- (TM-) polarized modes [(0,0),
(0,1), (1,0), (1,1), and (0,2)] in the fundamental field inside
the waveguide (see Fig.~\ref{fig:res}). As expected, the spatial
modal structure of the SH field has been found richer owing to
doubled relative waveguide dimensions with respect to the SH
wavelength. The experimental spatial intensity profiles shown in
Fig.~\ref{fig:res} are partially distorted in comparison with the
theoretical ones. This is caused by waveguide imperfections,
non-ideal coupling of the fundamental beam leading to the
excitation of more than one spatial mode and subsequent SHG into
several spatial modes. 

\begin{figure} 
    \includegraphics[width=\textwidth]{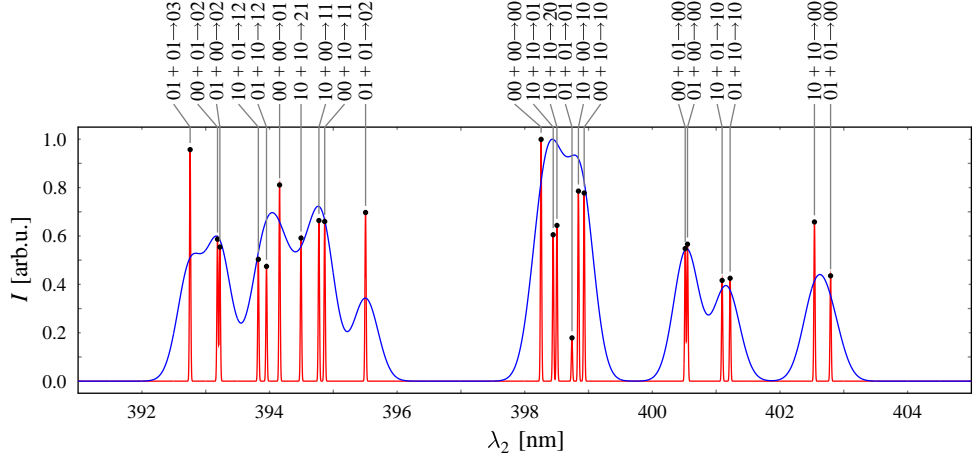}
\caption{Theoretical SH spectral intensity $I$ observed in an
ideal (red curve) and real (blue curve) structure. Individual
combinations of spatial modes are identified.} \label{fig:spec}
\end{figure}

In the nonlinear process, a sufficient overlap of three chosen
spatial mode profiles is needed to observe an individual nonlinear
process in the measured SH spectrum. As also phase-matching
conditions play an important role, a given individual process
occurs only in a restricted spectral range of SH field (see
Fig.~\ref{fig:res}). If spatial modes of the fundamental field are
uniformly excited by a spectrally broad pumping, several
individual nonlinear processes typically occur in the waveguide
and contribute to the SH spectrum. As a consequence, the expected
SH intensity spectrum is relatively broad (see
Fig.~\ref{fig:spec}). As SH spectra belonging to individual
processes are relatively wide and close to each other, only
spectral filtering of the SH field itself is not sufficient to
distinguish them. After spectral filtering, individual processes
become experimentally available by exciting only specific spatial
modes in the fundamental beam. Technically, the proper excitation
of the chosen spatial mode(s) is accomplished by suitable coupling
of the pump beam into the waveguide.

\begin{table} 
\centering
{\small
\begin{tabular}{c c c c c c}
\hline \hline
    Modes & $\lambda_{\rm 2, exp}$ & $\lambda_{\rm 2, theor}$
    & $\Delta\lambda_{\rm 2, exp}$ & $P_{\rm exp}$ & $P_{\rm theor}$ \\
    (TE + TM $\rightarrow$ TE) & [nm] & [nm] & [nm] & [arb.u.] & [arb.u.] \\
\hline
    00 + 00 $\rightarrow$ 00 & 398.3 & 398.6 & 0.8 & 1.00 & 1.00 \\
    00 + 00 $\rightarrow$ 01 & 394.2 & 394.4 & 0.9 & 0.65 & 0.81 \\
    10 + 00 $\rightarrow$ 10 & 398.8 & 399.5 & 0.6 & 0.42 & 0.78 \\
    01 + 01 $\rightarrow$ 00 & 402.8 & 403.0 & 0.8 & 0.38 & 0.43 \\
    10 + 01 $\rightarrow$ 10 & 401.1 & 401.9 & 0.9 & 0.35 & 0.42 \\
    01 + 01 $\rightarrow$ 01 & 398.7 & 397.1 & 0.6 & 0.32 & 0.18 \\
\hline \hline
\end{tabular}
} \caption{Individual SHG processes identified in the waveguide,
experimental central SH wavelength $\lambda_{\rm 2, exp} $,
theoretical central SH wavelength $\lambda_{\rm 2, theor} $,
spectral width $\Delta\lambda_{\rm{2, }exp} $, experimental
relative power $P_{\rm exp} $, and theoretical relative power
$P_{\rm theor} $ are given. Experimental central wavelengths are
measured with the error 0.1~nm given by the spectrometer response
function. Spectral widths (relative powers) have the relative
error 5\% (9\%).} \label{tab2}
\end{table}
Careful alignment of the pump-beam coupling and use of 3~nm wide
spectral filters in this beam have allowed to observe cca 10
spectral lines from the SH spectrum plotted in
Fig.~\ref{fig:spec}. SH central wavelengths $\lambda_{2}$,
spectral widths $\Delta\lambda_{2}$ as well as relative powers of
the observed SHG processes are summarized in Table~\ref{tab2}. The
comparison of theoretical and experimental parameters in
Table~\ref{tab2} has revealed good agreement in central
wavelengths of the SH lines. Larger differences between the
theoretical and experimental relative powers $P$ can be explained
by the difficulties in reaching an efficient coupling of the
corresponding pump mode. The experimental spatial intensity
profiles of the three interacting modes and the corresponding SH
intensity spectra are plotted in Fig.~\ref{fig:res} for three
typical mode combinations. These graphs document real capabilities
in controlling spatial properties of the SH beam. If the analyzed
waveguide were used for parametric down-conversion, several
individual processes based on different spatial modes would
simultaneously occur contributing together to the signal and idler
fields' profiles [e.g., parametric processes (0,0) $\rightarrow$
(0,0) + (0,0) and (0,0) $\rightarrow$ (0,1) + (0,1)]. This might
be used for the generation of modally entangled photon pairs
\cite{Saleh2009}. We finally note that temperature tuning is
important in the determination of efficient individual processes.
This gives additional possibilities in controlling the properties
of the generated fields.

\section{Conclusions}

The process of second harmonic generation in a periodically-poled
KTP waveguide has been investigated both experimentally and
theoretically. Three types of the nonlinear processes (type~II, I,
and 0) have been observed simultaneously utilizing the first,
second and third harmonics of the spatial nonlinear modulation. A
scalar finite elements method based on the Galerkin method has
been applied to calculate spatial mode profiles, propagation
constants, and frequencies of the interacting fields. It has been
shown that positions of the spectral lines of individual processes
depend strongly on waveguide's depth and width in ideal
waveguides. In real waveguides, fabrication imperfections
considerably broaden single spectral lines occurring in ideal
waveguides. Individual spatial and spectral processes in type~II
interaction have been experimentally characterized by changing
pump-beam spectral filtering and coupling into the waveguide. The
analyzed waveguide has been recognized as a versatile source of
second-harmonic light.

\section*{Acknowledgements}

Support by the project IAA100100713 of GA AS CR is acknowledged.
The authors gratefully acknowledge the support by the Operational
Program Research and Development for Innovations - European
Regional Development Fund project CZ.1.05/2.1.00/03.0058 and
project COST OC09026 of the Ministry of Education. R.M.
acknowledges the support by the project PrF\_2012\_003 of
Palack\'{y} University. J.Sv. thanks the project FI-DGR 2011 of
The Catalan Government. R.M. and J.Sv. also gratefully acknowledge
the support by the Operational Program Education for
Competitiveness - European Social Fund (project
CZ.1.07/2.3.00/20.0017 of the Ministry of Education).

\end{document}